# AI and MBTI: A Synergistic Framework for Enhanced Team Dynamics

Wang Yue

Singapore University of Social Sciences

wangyue@suss.edu.sg

https://orcid.org/0000-0001-8322-5384

**Abstract:** This paper proposes a theoretical framework for understanding and leveraging the synergy between artificial intelligence (AI) and personality types as defined by the Myers-Briggs Type Indicator (MBTI) in organizational team settings. We argue that AI capabilities can complement and enhance different MBTI types, leading to improved team performance. The AI-MBTI Synergy Framework is introduced, focusing on the Intuition-Sensing and Thinking-Feeling dimensions. We present propositions about how AI can augment team dynamics across four team types: Visionary, Strategic, Supportive, and Operational. A novel implementation is proposed to create an intelligent team optimization algorithm. Implications for theory and practice are discussed, along with directions for future research.

*\* Conflict of Interests Statement*

*The authors declare that there are no conflicts of interest regarding the publication of this paper. All research was conducted independently, without any financial or personal relationships that could have influenced the work reported in this paper. The funding sources for this research did not have any involvement in the study design, data collection, analysis, interpretation, writing, or decision to submit the article for publication. The authors have no affiliations with or involvement in any organization or entity with any financial interest, or non-financial interest in the subject matter or materials discussed in this manuscript.*

*\*Data Availability Statement*

*No data were generated or analyzed in this study. Consequently, there are no data sets associated with this paper. The research and conclusions presented are based on theoretical analysis and literature review, and therefore do not require a data availability statement.*

# 1. Introduction

As artificial intelligence (AI) becomes increasingly integrated into organizational processes, understanding its interaction with human factors is crucial for optimizing team performance (Braz & Sichman, 2020). This paper explores the synergistic relationship between AI and personality types as defined by the Myers-Briggs Type Indicator (MBTI), with a focus on enhancing team dynamics across different personality combinations.

The MBTI, developed by Myers and Briggs based on Jung (1971)'s theory of psychological types, has been widely used in organizational settings to understand individual differences and improve team composition (Myers et al., 1998). Its applications range from recruitment and job matching (Chen & Chen, 2015) to team building and conflict resolution (Bradley & Hebert, 1997; Wilsher, 2015). However, research on how AI capabilities might complement or enhance different MBTI types remains limited. This study aims to address this gap by proposing a theoretical framework for AI-MBTI synergy in team contexts.

Drawing on cognitive process theory and recent advances in AI applications, we develop propositions about how AI can augment the strengths and mitigate the weaknesses of different MBTI-based team compositions. Our framework focuses primarily on the Intuition-Sensing and Thinking-Feeling dimensions, as these most directly relate to information processing and decision-making - areas where AI can provide significant enhancement (Salvit & Sklar, 2010).

The Belief-Desire-Intention (BDI) model, a foundation for rational agents in AI, provides a starting point for integrating MBTI into agent behavior (Rao & Georgeff, 1995). Salvit and Sklar (2012) extended this model to incorporate MBTI personality types, demonstrating how different types influence agent decision-making and performance in simple tasks. Building on this work, we propose to further extend the BDI model to study more complex team dynamics in organizational settings.

The application of deep learning techniques to personality detection has shown promising results. Lynn et al. (2020) proposed a hierarchical modeling approach that incorporates message-level attention, improving personality prediction accuracy. Verhoeven et al. (2016) introduced TwiSty, a multilingual Twitter corpus for personality profiling, opening avenues for cross-cultural personality research. These advancements, as reviewed by Mehta et al. (2020), demonstrate the growing sophistication of AI approaches in personality computing.

By leveraging multiagent systems (MAS) and the extended BDI model, we can simulate and analyze how teams with diverse MBTI compositions interact and perform in various scenarios (Pereira et al., 2005; Padgham & Lambrix, 2000). This approach allows us to explore how AI can be tailored to complement different personality types, potentially leading to more effective human-AI collaboration and higher-performing teams.

Our research contributes to the growing field of personality computing in AI and offers practical insights for organizations seeking to optimize their human-AI teams. By understanding the interplay between MBTI types and AI capabilities, we can develop more nuanced and effective strategies for team formation, task allocation, and overall organizational performance (Maxon, 1985; McClure & Werther, 1993).

In this paper, we not only present a theoretical framework but also propose a novel implementation of an AI-powered team assistant that integrates MBTI insights with state-of-the-art LLM. This implementation aims to provide a practical tool for organizations to optimize team composition and enhance team dynamics based on the AI-MBTI Synergy Framework.

## 2. Theoretical Background

The Myers-Briggs Type Indicator (MBTI) has been a prominent tool in organizational psychology for decades, offering insights into individual differences and team dynamics. Rooted in Jung's theory of psychological types, MBTI categorizes individuals along four dimensions: Extraversion-Introversion, Sensing-Intuition, Thinking-Feeling, and Judging-Perceiving (Myers et al., 1998). Recent developments in MBTI research have expanded its applications and refined its methodologies (Gu & Hu, 2012).

Cognitive processes in MBTI theory encompass eight functions derived from Jung (1971)'s work on psychological types. These include four perception processes (Extraverted and Introverted Sensing and Intuition) and four judgment processes (Extraverted and Introverted Thinking and Feeling). These processes describe how individuals gather information and make decisions. While not cognitive processes themselves, the J (Judging) and P (Perceiving) preferences in MBTI indicate how these cognitive functions are expressed externally. For extraverts, J indicates a dominant judging function (Thinking or Feeling), while P indicates a dominant perceiving function (Sensing or Intuition). For introverts, J and P refer to the auxiliary function used in the external world. The J/P dichotomy thus helps determine the order and expression of cognitive functions in each MBTI type, influencing how individuals interact with their environment and which cognitive processes they tend to use more visibly.

In organizational settings, MBTI has been applied to various aspects of human resource management. Chen and Chen (2015) explored the mechanism of MBTI personality types and job matching, highlighting the importance of aligning individual preferences with job requirements. Wilsher (2015) discussed the implications of behavior profiling for recruitment and team building, emphasizing the potential of MBTI in creating more cohesive and effective teams. Yang (2022) explored the application of MBTI in organizational settings, discussing its potential benefits for team building, communication improvement, and employee development within companies. Bradley and Hebert (1997) investigated the effect of personality type on team performance, finding that balanced teams with diverse MBTI types often outperform homogeneous teams. Maxon (1985) and McClure and Werther (1993) further explored the role of personality in personal development and management interventions, suggesting that MBTI-informed approaches can enhance leadership and team effectiveness.

The integration of MBTI with artificial intelligence and computational methods has opened new avenues for research and application. Salvit and Sklar (2010) proposed a model for agent behavior in multiagent teams based on MBTI, laying the groundwork for simulating personality-driven interactions in AI systems. This work builds on the Belief-Desire-Intention (BDI) model for rational agents (Rao & Georgeff, 1995), which provides a framework for representing and reasoning about goals and intentions in AI systems.

Recent advancements in natural language processing and machine learning have enabled automated personality prediction from text data. Gjurković and Šnajder (2018) demonstrated the potential of using Reddit data for MBTI prediction, while Plank and Hovy (2015) achieved similar results using Twitter data. These studies highlight the rich potential of social media data for personality computing. The PANDORA dataset, introduced by Gjurković et al. (2021), offers a comprehensive resource for studying multiple personality models, including MBTI, across diverse demographic groups.

Lynn et al. (2020) proposed a hierarchical modeling approach for personality prediction, incorporating message-level attention to improve accuracy. This work demonstrates the potential of deep learning techniques in capturing nuanced personality traits from text data. Verhoeven et al. (2016) contributed to the field by introducing TwiSty, a multilingual Twitter corpus for personality profiling, enabling cross-cultural studies of personality expression in social media. Cerkez et al. (2021) proposed a novel method for MBTI classification using machine learning techniques, focusing on the impact of class components to improve accuracy in personality type prediction. The application of deep learning to personality detection has seen significant growth, as reviewed by Mehta et al. (2020). These techniques offer promising avenues for automating MBTI assessment and integrating personality insights into AI-driven decision-making systems.

By combining these computational approaches with traditional MBTI theory and organizational psychology, we can develop more sophisticated models of human-AI interaction in workplace settings. This interdisciplinary approach holds the potential to enhance team formation, improve communication, and optimize task allocation in mixed human-AI teams.

The theoretical background presented here sets the stage for our proposed framework, which aims to leverage AI capabilities to complement and enhance MBTI-based team dynamics. By bridging the gap between personality psychology and artificial intelligence, we seek to create more effective, harmonious, and productive organizational environments that capitalize on the strengths of both human and artificial team members.

## 3. Conceptual Framework: The AI-MBTI Synergy Model

### 3.1 AI's MBTI Alignment

P1: AI exhibits characteristics most similar to the INTJ personality type, emphasizing introverted intuition (Ni) for pattern recognition and extraverted thinking (Te) for logical analysis and decision-making.

This proposition is grounded in the cognitive processes theory of MBTI. Introverted Intuition (Ni) is characterized by an ability to recognize patterns and implications in complex data, which aligns closely with AI's capacity for deep learning and pattern recognition (Mehta et al., 2020). Extraverted Thinking (Te) involves logical analysis and decision-making based on objective criteria, mirroring AI's data-driven approach to problem-solving.

In organizational contexts, INTJ types are often associated with strategic planning and systems thinking (Bradley & Hebert, 1997). Similarly, AI systems excel at analyzing complex systems and providing strategic insights, making them valuable tools for organizational decision-making. This alignment suggests that AI can be particularly effective in augmenting roles and tasks typically suited to INTJ personalities.

3.2 MBTI Type Most Complemented by AI

P2: The ENFP personality type is most complemented by AI's capabilities, creating a synergistic relationship that enhances team performance.

ENFPs, characterized by dominant Extraverted Intuition (Ne) and auxiliary Introverted Feeling (Fi), are known for their creativity, enthusiasm, and people-oriented approach (Myers et al., 1998). However, they may struggle with details and logical analysis. AI's strengths in data processing and logical reasoning can complement these potential weaknesses.

In organizational settings, ENFPs often excel in roles requiring innovation and interpersonal skills (Chen & Chen, 2015). By pairing ENFPs with AI systems, organizations can create a powerful synergy: the ENFP's creative ideation and emotional intelligence combined with AI's analytical capabilities. This combination could be particularly effective in areas such as product development, marketing, and customer relations.

3.3 The AI-MBTI Synergy Framework

P3: AI enhancement will significantly improve team performance by complementing the dominant cognitive processes of each team type.

This proposition builds on research showing that diverse teams often outperform homogeneous ones (Bradley & Hebert, 1997). By tailoring AI support to each team type's cognitive preferences, we can enhance their strengths while mitigating weaknesses.

Our framework identifies four team types based on the N/S and T/F combinations:

  a. Visionary Teams (NF)
  b. Strategic Teams (NT)
  c. Supportive Teams (SF)
  d. Operational Teams (ST)

For example, Visionary Teams (NF) excel at generating ideas and understanding people but may struggle with implementation. AI can provide data-driven validation of ideas and help translate visions into actionable plans. Strategic Teams (NT) benefit from AI's advanced analytical capabilities, enhancing their ability to make complex decisions. Supportive Teams (SF) can leverage AI for personalized customer interactions, while Operational Teams (ST) can use AI for process optimization and quality control. Please refer to the below Figure 1 for more detailed description of the four team's settings.

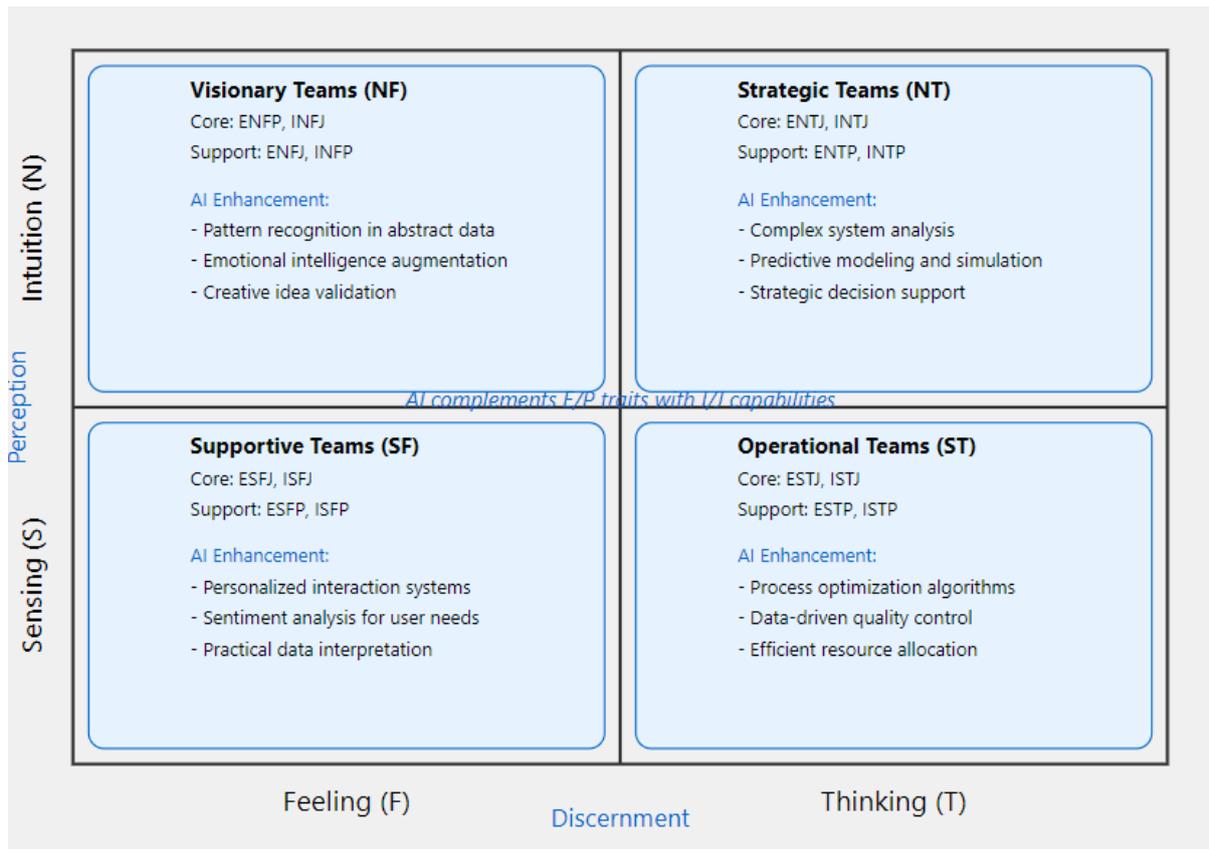

Figure 1: AI-MBTI Team Synergy Framework

This framework aligns with Yang's (2022) findings on the application of MBTI in organizations, suggesting that understanding team composition can lead to more effective collaboration and improved performance.

3.4 AI-Enhanced Cognitive Processes

P4: AI systems can augment and enhance the eight cognitive processes (Se, Si, Ne, Ni, Te, Ti, Fe, Fi) associated with MBTI types, leading to improved decision-making and problem-solving in teams.

This proposition extends the BDI model (Rao & Georgeff, 1995) and its MBTI adaptations (Salvit & Sklar, 2012) to include AI augmentation of cognitive processes. For instance, AI can enhance Extraverted Sensing (Se) through advanced sensor technologies, Introverted Intuition (Ni) through predictive analytics, and Extraverted Feeling (Fe) through sentiment analysis and emotional intelligence algorithms.

In organizational contexts, this enhancement could lead to more balanced decision-making. For example, a leader with strong Ti (Introverted Thinking) could use AI to augment their Fe, helping them better understand and respond to team emotions. This aligns with McClure and Werther's (1993) work on personality variables in management development, suggesting that such augmentation could lead to more effective leadership.

3.5 Dynamic Team Composition

P5: AI-driven personality assessment and team formation algorithms can dynamically optimize team composition based on task requirements and individual MBTI profiles, leading to improved overall team performance.

This proposition leverages recent advancements in AI-based personality prediction (Lynn et al., 2020; Gjurković et al., 2021) and applies them to team formation. By continuously assessing personality traits through natural language processing of workplace communications (Plank & Hovy, 2015), AI can suggest optimal team compositions for specific tasks.

This approach aligns with Wilsher's (2015) insights on behavior profiling for team building, but takes it a step further by making it dynamic and data-driven. It also addresses the challenge of team diversity noted by Bradley and Hebert (1997), ensuring a balance of personality types suited to each task.

In practice, this could mean AI systems recommending team reshuffles for different project phases, ensuring the right mix of visionaries, strategists, supporters, and operators at each stage. This dynamic approach to team composition could lead to more agile and effective organizations, better equipped to handle diverse challenges.

By integrating insights from MBTI theory, cognitive processes, organizational psychology, and state-of-the-art AI capabilities, this expanded framework provides a comprehensive approach to enhancing team performance through AI-MBTI synergy.

## 4. Proposed Implementation

To translate our theoretical framework into a practical tool, we propose a novel implementation of an AI-powered team optimization algorithm that integrates MBTI insights with advanced AI technologies. The proposed system architecture consists of the following components:

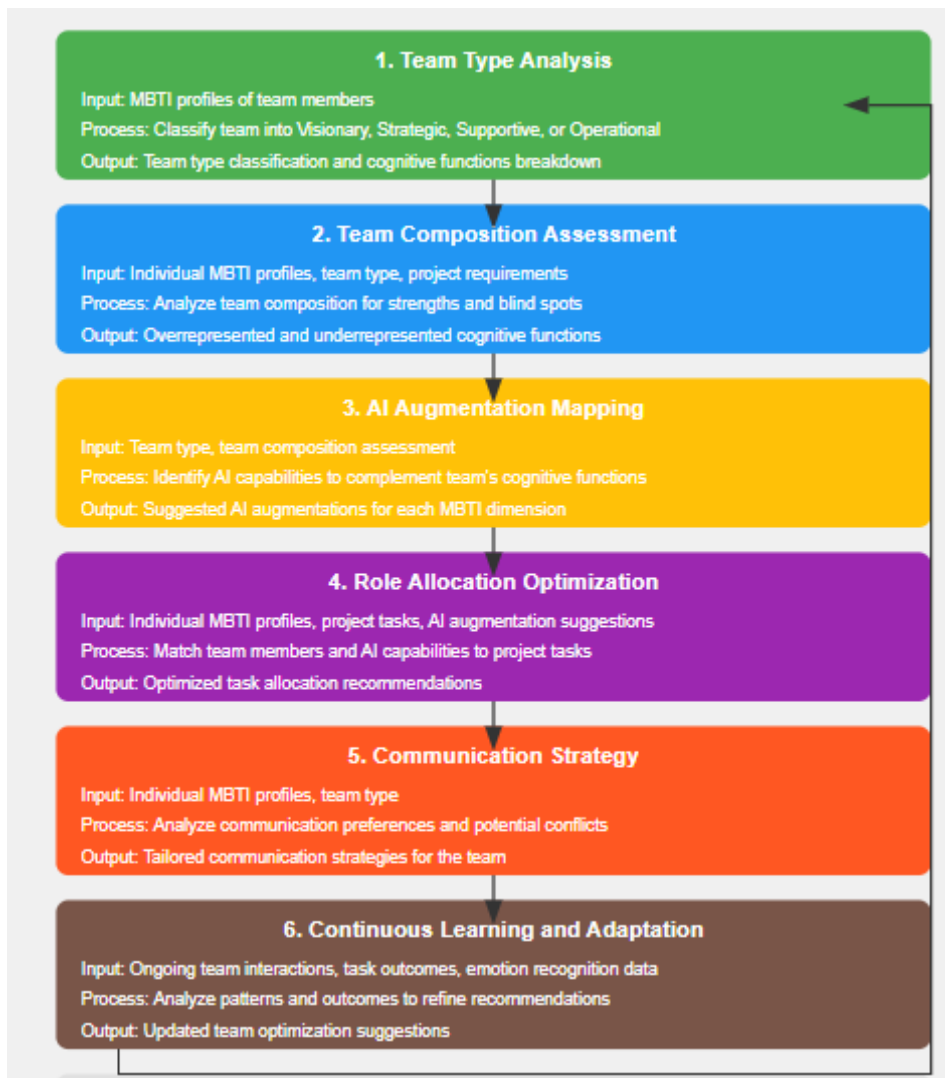

Figure 2: Team Optimization Algorithm

The Team Optimization Algorithm shown in Figure 2 aims to enhance team performance by leveraging MBTI insights and AI capabilities. Here is a more detailed breakdown of its components and functionality:

1) The process begins with the input of MBTI profiles of team members. The algorithm classifies the team into one of the four types (Visionary, Strategic, Supportive, Operational) based on the dominant cognitive functions present. The output of this stage is a team type classification and a breakdown of the cognitive functions represented within the team.
2) Next, the algorithm takes individual MBTI profiles, the identified team type, and project requirements as inputs. It then analyzes the current team composition to identify strengths and potential blind spots. The output of this assessment is the identification of overrepresented and underrepresented cognitive functions within the team.
3) In this next step, the algorithm uses the team type and composition assessment to identify AI capabilities that can complement the team's cognitive functions. The output is a set of suggested AI augmentations tailored for each MBTI dimension, ensuring that the team's strengths are enhanced, and weaknesses are mitigated.

4) With individual MBTI profiles, project tasks, and AI augmentation suggestions as inputs, the algorithm matches team members and AI capabilities to project tasks based on cognitive preferences. The output is an optimized task allocation recommendation that aligns team members' natural strengths with the tasks at hand, supported by AI where necessary.
5) The algorithm analyzes the communication preferences and potential conflicts within the team, using the individual MBTI profiles and team type as inputs. The output is a tailored communication strategy designed to facilitate effective interactions and reduce misunderstandings within the team.
6) Finally, the algorithm monitors ongoing team interactions, task outcomes, and emotion recognition data. It analyzes these patterns and outcomes to refine its recommendations continually. The output is updated team optimization suggestions that adapt to the evolving dynamics and needs of the team.

Consider a Visionary Team (NF dominant) working on a new product development project. The team comprises three ENFPs, two INFJs, and one ENTJ.

1) The input consists of the MBTI profiles of the team members. The output identifies the team as a Visionary Team (NF dominant), strong in ideation and big-picture thinking but potentially weak in detailed execution.
2) The assessment reveals that the team has overrepresented cognitive functions in Intuition (N) and Feeling (F), while Sensing (S) and Thinking (T) are underrepresented.
3) To address these imbalances, the algorithm suggests augmenting Sensing (S) with AI-driven data analysis and market research, and augmenting Thinking (T) with AI-powered project management and decision-making tools.
4) Based on the MBTI profiles and AI augmentation suggestions, the algorithm recommends that ENFPs focus on brainstorming and customer experience design, supported by AI market research. INFJs should handle long-term strategy and team harmony, aided by AI trend analysis. The ENTJ should take on project leadership, leveraging AI project management tools.
5) The algorithm suggests implementing a balanced approach of brainstorming sessions for N types and structured updates for J types. It also recommends using AI to summarize discussions to accommodate different communication preferences.
6) The algorithm monitors team interactions and project milestones, suggesting adjustments such as increasing structured check-ins if deadlines are missed or recommending AI-driven prototyping tools if execution is lagging.

By integrating these components, the Team Optimization Algorithm continuously balances the team's natural strengths with AI augmentation, addressing potential weaknesses and optimizing overall performance. This approach creates a dynamic, responsive system that can adapt to changing project needs and team dynamics over time.

## 5. Discussion

5.1 Implications for Theory

The AI-MBTI Synergy Framework contributes significantly to the existing literature on MBTI, AI, and team dynamics in several ways. First, it integrates AI and MBTI, bridging the gap between personality psychology and artificial intelligence. This integration offers a novel perspective on how AI can complement and enhance human cognitive processes, extending the work of Salvit and Sklar (2010, 2012) by applying MBTI-based agent modeling to real-world organizational contexts.

Second, by incorporating MBTI cognitive processes into the BDI model (Rao & Georgeff, 1995), our framework provides a more nuanced understanding of agent behavior in multi-agent systems. This expansion aligns with recent trends in personality computing (Mehta et al., 2020) and offers new avenues for research in AI-augmented decision-making.

Third, the proposed framework builds upon existing research on MBTI in team formation (Bradley & Hebert, 1997; Wilsher, 2015) by introducing AI-driven, dynamic team composition. This approach offers a more adaptive and responsive model of team building that can respond to changing task requirements in real-time.

Lastly, by combining insights from organizational psychology, personality theory, and artificial intelligence, this framework contributes to the growing field of computational personality recognition (Gjurković & Šnajder, 2018; Lynn et al., 2020). It offers a more holistic approach to understanding and enhancing team dynamics.

The proposed implementation extends the theoretical framework by providing a concrete example of how AI and MBTI insights can be integrated into a practical tool. This bridges the gap between theory and practice, offering new avenues for research on the real-world applications of AI-enhanced personality assessments in organizational settings.

5.2 Implications for Practice

The AI-MBTI Synergy Framework has several practical applications for organizations. HR professionals and team leaders can use AI-powered tools based on this framework to create more balanced and effective teams, considering both personality types and task requirements. Organizations can develop AI systems tailored to complement specific MBTI types, providing personalized support that enhances individual and team performance.

Moreover, the framework can inform leadership development programs, helping leaders understand how to leverage AI to compensate for their cognitive biases and enhance their decision-making processes. By understanding the interplay between AI capabilities and MBTI types, organizations can develop more effective strategies for managing and resolving conflicts within teams. The dynamic team composition aspect of the framework can help organizations allocate tasks more efficiently, matching project requirements with the most suitable combination of human and AI capabilities.

The AI-powered team optimization algorithm offers organizations a sophisticated tool for real-time team optimization. By providing personalized insights and recommendations based on MBTI profiles, the system can help managers make more informed decisions about team composition, task allocation, and conflict resolution.

5.3 Limitations and Future Research Directions

While the AI-MBTI Synergy Framework offers promising insights, several limitations should be acknowledged. The propositions presented in this paper require rigorous empirical testing.

Future research should focus on developing experiments and field studies to validate the framework's predictions.

Additionally, the current framework is primarily based on Western conceptualizations of personality. Future research should explore how cultural differences might impact the AI-MBTI synergy, perhaps building on cross-cultural datasets like TwiSty (Verhoeven et al., 2016). The use of AI for personality assessment and team composition raises important ethical questions about privacy, consent, and potential biases. Future research should address these ethical considerations.

The long-term impact of AI-augmented team dynamics on individual well-being and organizational culture needs to be studied. While this framework focuses on MBTI, future research could explore how it might be extended or compared to other personality models, such as the Big Five.

Future research directions could include developing and testing AI algorithms for real-time personality assessment and team optimization based on the framework. Investigating the impact of AI-MBTI synergy on specific organizational outcomes, such as innovation, productivity, and employee satisfaction, and exploring how the framework might be applied in different industries and organizational structures. Additionally, examining the potential for AI to help individuals develop their non-dominant cognitive functions, leading to more balanced personality types, is another area of interest.

## 5. Conclusion

The AI-MBTI Synergy Framework and its proposed implementation offer a novel approach to understanding and enhancing team dynamics in the age of AI. By leveraging AI capabilities to complement different MBTI types, organizations can create more balanced, efficient, and innovative teams. The practical implementation of these ideas through an AI-powered team assistant demonstrates the potential for AI to revolutionize team management and organizational behavior. As AI continues to evolve, frameworks and tools like these will be crucial in ensuring that we harness the full potential of both human and artificial intelligence in organizational settings.